# Public vs Media Opinion on Robots

Alireza Javaheri* (Independent Researcher), Navid Moghadamnejad* (Independent Researcher), Hamidreza Keshavarz (Tarbiat Modares University), Ehsan Javaheri (The Technical University of Berlin), Chelsea Dobbins (The University of Queensland), Elaheh Momeni (University of Vienna), Reza Rawassizadeh (Boston University)

* These authors have contributed equally to the paper and they are co-first authors

Fast proliferation of robots in people's everyday lives during recent years calls for a profound examination of public consensus, which is the ultimate determinant of the future of this industry. This paper investigates text corpora, consisting of posts in Twitter, Google News, Bing News, and Kickstarter, over an 8-year period to quantify the public and media opinion about this emerging technology. Results demonstrate that the news platforms and the public take an overall positive position on robots. However, there is a deviation between news coverage and people's attitude. Among various robot types, sex robots raise the fiercest debate. Besides, our evaluation reveals that the public and news media conceptualization of robotics has altered over the recent years. More specifically, a shift from the solely industrial-purposed machines, towards more social, assistive, and multi-purpose gadgets is visible.

## 1   INTRODUCTION

The fourth industrial revolution is most likely to alter the nature of work, business, and society in the coming decades [1, 2, 3, 4]. Not only are fast developments spurring in individuals' daily routines, but also fresh potentials are being made in the market [3]; through which, new enterprises like iRobot[1] and Touch Bionics[2] are emerging and demonstrating success. Microsoft co-founder, William Henry Gates III predicted that robots would inevitably be on the rise, just as personal computers were in the end of the last century [5]. Robotic surgery [6], controlling autonomous vehicles [7], and tutoring students [8] are among the responsibilities that are being assumed by robots, and yet there would likely be more developments [9]. On the other hand, the accelerating growth of robot capabilities have raised concerns, for example, a substantial worry is that they might be a substitute for human labor [10, 11], be employed as means of warfare [12], or as Stephen Hawking warned, spell the end of the human race [13].

Amid such rapid proliferation, the debate over their development, which is getting fiercer, underscores the need for extensive research in order to determine the public's attitudes towards them. To our knowledge, the number of such studies is scarce. As an attempt to bridge this gap, it is essential to explore the following questions:

RQ1) What is the public's opinion about robots in general and whether news and media support this opinion?
RQ2) What type of robots do the public prefer or dislike most?
RQ3) How does news and public opinion about robots evolve over time?

The first question focuses on a general consensus, since people are the ultimate decision makers who will shape the future of the robotics industry. On the other hand, ignoring the position of news media on the same topic would be naive. Considering the pivotal role of media in society, and its ability to lead or sway public opinion [14, 15, 16], it is crucial to identify whether media takes a position similar to the public on robots or not. The second question attempts to acquire in-depth knowledge of the public's preferences for robots. Given the broad range of robot types, exploring the popularity of each type is also necessary. The third question deals with the evolution of news and public approach toward various robot types, over the period of time during which the study was conducted, in order to gain an understanding of the trend in recent years.

---

[1] https://www.irobot.com/
[2] http://touchbionics.com/

To answer these three questions, we investigate Twitter as a social media platform, which is growing in popularity and serves as a repository of billions of attitude expressions. Exploring a general consensus of a wide spectrum of topics in Twitter is the focus of many studies [17, 18, 19, 20]. Various works show the advantages of social media mining [21, 22] and illustrate that the data found on social media can compare with the data collected otherwise. Besides, tweets are immune from the typical errors inherent in traditional means of information gathering (e.g. polls and questionnaires) [23, 24, 25], where the participants' attitudes are contingent upon the context of the questions, their format, wording, and ordering.

In the first step of this paper, recent posts from Twitter (from mid-2018, as the platform does not allow for longitudinal data to be collected), and longitudinal data from Google News, Bing News and Kickstarter in the period between 2011 and 2018 are extracted using a specially devised algorithm. This corpus allows us to implement clustering to contrast attitudes of people against News. In the next step, a qualitative investigation on tweets is conducted by annotating 180 tweets by two independent raters. The final step focuses on topic evolution of news and Kickstarter articles by means of an algorithm based on word clustering [26].

Results demonstrate that there is a noticeable deviation between News coverage and public opinion on robots. While both share an overall positive position, news is more conservative and involved with non-polarized posts. Among various robot types, sex robots raise the fiercest debate, however their proponents are about twice as many as the opponents. Social robots and service robots also are highly popular. And finally, the evolution study observes that the public and news media's conceptualization of robotics has altered over recent years, and shifted from solely industrial-purpose machines, towards more in-home, friendly, social, and multi-purpose gadget.

People's view is of paramount importance, since they ultimately shape demand and regulation in certain areas, by urging companies and legislators to restrict or expand research in those areas. In response to such worries, scientists can alter the direction of their research in order to alleviate concerns. Results of our study can help investors better recognize the potentially successful or unsafe areas in robotic industries, guide robot developers in the early stages of their product development to adapt them to the users' demands, cast light on features of most popularity for engineers to focus on, and finally offer marketers with products, which fulfill more market expectations.

## 2 RELATED WORKS

Technological advancements entangle our lives, including robots, artificially intelligent systems and interfaces. The future of robots is unknown [27], but what we think of them right now, might be able to re-shape it. The public's acceptance or rejection of them can result in altering investments in the industry. Therefore, it is important to understand the concerns that people have, regarding daily operation of, or collaborations with robots, and what they expect from them.

### 2.1 Public Opinion about Future Technological Trends

Some of humankind's biggest concerns, such as global warming and climate change [28], poverty [29], terrorism [30], political debates [31], and privacy issues [32] have been researched and peoples' perspectives on these subjects have been studied through social media and opinion mining. However, regarding the subject of robots, a mere of inadequate research has been conducted. A promising number of investigations have been conducted on the influences of artificial intelligence (AI). Public perception of AI and the future of humanity affected by AI has been studied [33] and the results show that the public has a generally positive view on AI, while the experts are more conservative. Studies of the future progress of artificial intelligence [34] from the viewpoint of AI experts predicted that the future belongs to the intelligent and super intelligent machines. It also predicted a chance of one in three that this development will have negative consequences for humanity. For instance, the survey carried out on AI acceptance [35] indicates that consumers will most likely feel the impact of AI within the next five years, and the prospect of such impact is met with mixed emotions, with 60% having a positive opinion. Another survey [36] shows that consumers are concerned about job loss, security issues, and privacy infringement, but for the most part, they are accepting of AI. About 45% of all participants believe the effects of AI on society is positive. In an effort to descend further back into the history of AI and the trends of public views on it [37], public perception of AI has been studied through articles in the New York Times over a 30-year period. Similarly, the results reveal that the public views on AI have stayed optimistic over time, while AI itself has taken a stronger role in public discussions, especially in the field of healthcare. Another study [38] offers insight into the fact that media coverage of armed conflicts has raised public attention towards these conflicts. A parallel study [39] reveals the positive influence of certain literature used in news articles and the popularity of the message they convey.

A study of the public's fears towards robots [40] shows people tend to think of robots as harmful entities that will challenge specificity in western cultures. In contrast, the Asian (Japanese) public do not seem to think that robots will affect the human specificity. Although useful, these comparisons make definitive statements about specific cultures and regions. Alternatively, another study [41] demonstrates that the average European citizen, thinks very positively of robots, especially in manufacturing and security sectors, but not so much in areas such as healthcare and child care. In a small-scale study of public expectations and fears towards the future of robots [42], people display rather negative judgements towards these machines, especially as future role models for their children, though they are more open to their utilization for public security.

## 2.2 Human-Robot Interactions (HRI)

As helpful as they might be in our future lives, over-reliance on intelligent robots might hurt humans. Thus, the opinions of people on the present and future roles of these smart machines and interactions with them needs to be thoroughly researched.

A study on cross-cultural acceptance of tutoring robots [43] reveals that in Korea and Japan, while people are more open to purchasing tutoring robots, they have higher expectations of these robots. In contrast, in Spain people are more negative towards acquiring such robots for learning purposes. Human-robot interactions in the Middle East is studied [44] through utilizing a humanoid robot in a public mall. Amongst the findings of this work is that people from the Middle East had more favorable views towards humanoid robots than people from North Africa. Although interestingly, these results are also restricted to a specific region.

In a different study on negative attitudes towards communication robots in particular [45], individuals' ability to accept such robots is associated with their emotions towards those robots. Their contribution can help designers predict the acceptance of their robots. However, the study is limited to a group of 38 Japanese participants. In a study on the interactions of the elderly with instructor robots [46], while seniors moderately accept the robots and the administrators are enthusiastic about them, it is claimed that more work needs to be done on the auditory capability of the robots.

The increase in HRI applications has emphasized the demand for HRI ethics guidelines. Research [47] claims that ethics must be spread throughout and integrated from the beginning of design processes. As widely noticeable, most works in this field are mainly restricted to certain regions or have a rather limited number of participants. In this work, we have conducted a large-scale analysis on social media posts and news articles related to robots of different types and their temporal trends.

## 3 DATASET

As previously mentioned, the purpose of this work is to analyze both the public and news media points of view on robots. To this end, four different platforms were targeted for our web crawling. Two of which are news aggregation platforms (articles from Google News and Bing News), and the other two are public media platforms carrying public sentiments (records from Twitter and from Kickstarter). We have employed the eMentalist[3] API to collect web data. Specifically, for the Twitter data, the official API of Twitter, in addition to the technique of frequent keyword selection, are utilized to crawl the tweets from common Twitter users who have tweeted about robots. The majority of the datasets are shaped by records from 2011 to 2018. A total set of 238 keywords from different areas of robotics was created. This list made it possible to collect the related records in a database. In the database, text, timestamp of the records, and their platform are stored. The result was an initial dataset, which was then filtered (using a Python[4] code that eliminated the records not containing any keywords from the previous set) to ensure that the irrelevant entries were ruled out. This process culminated with the number of entries shown in Table 1.

Table 1. Number of Records in Datasets Before and After Filtering

| Platform | Initial Records | Final Records |
| --- | --- | --- |
| Twitter | 47,939 | 37,807 |
| Kickstarter | 2,000 | 1,901 |
| Google News | 6,400 | 5,451 |
| Bing News | 12,755 | 11,473 |

## 4 METHODS

Clustering is the technique being employed to group the datasets. It enables us to gain a factual summarization of the public's beliefs in the form of tweets, developers' expressions in the form of Kickstarter posts, and the news coverage through Google News and Bing News articles. In addition, we are able to compare and contrast the resulting clusters. Several well-known lexicons have been used to assign scores to each record. These scores will be treated as the features, which will later be used for clustering.

Lexicons is a highly used method to assess the positivity and/or negativity level of sentiments in texts, according to the words used in the text [48]. In comprehensive research carried out on lexicons [49], it is explained that each lexicon consists of a set of different words, and a balancing method of scoring the words. In this research, six lexicons were utilized (AFFIN, Liu Lexicon, NRC-emoticon, SenticNet, and SentiWordNet) in an effort to study the polarity of each record.

---

[3] http://ementalist.ai
[4] Python 3.7.0

To make use of lexicons, we devised a Python code to break each text record into separate words, look up the exact words in every lexicon, accumulate the corresponding lexicon scores of all the words of each record, and collect all the scores in a separate database for further processing. Fig. 1 illustrates the algorithm that calculates the AFFIN score generated for an arbitrary tweet record.

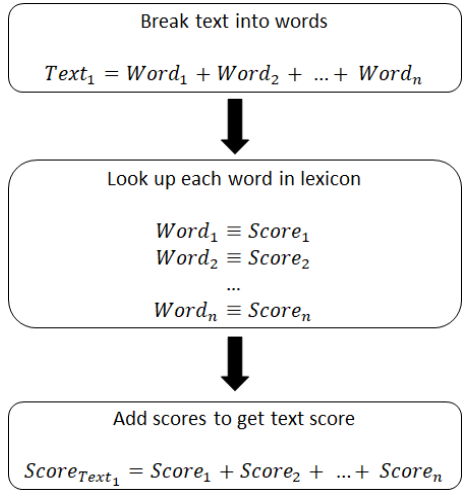

Fig. 1. A schematic diagram of the lexicon-based scoring algorithm for text entries

**Feature Selection:** For each database of *n* records, the algorithm is repeated 6*n times. Following running the algorithm for each of the six lexicons, over all the records, it is necessary to measure their overall positive and negative sentiments. This means that each of the six lexicons can be divided into two groups of positive words and negative words. As a result, each lexicon will yield two scores for each record:

1. The sum of all positive scores of each record
2. The sum of all negative scores of each record

The aggregate of these two scores will sum up the overall sentiment of each record. For example: "@BubbloApp: #AI is the Future - This happy robot helps kids with autism". This record receives a 0.384615 positive score, and 0 negative score from AFINN lexicon. Therefore, the overall score is positive, and it complies with the actual positive content of the tweet. As a result, a data matrix was produced, with 12 columns, each two of which belongs to one of the lexicons (see Fig. 2). An additional 13th column was added to the matrix to record each entry's database ID, for the purpose of tracking the data.

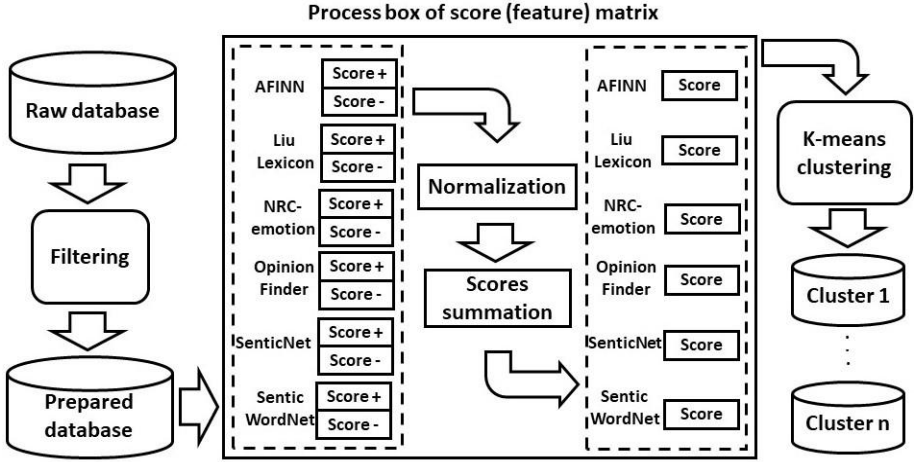

Fig. 2. Schematic diagram of clustering process

**Normalization:** The scores range varies from lexicon to lexicon. For instance, scores of each word in AFFIN lexicon range from -5 to 5, while these scores alternate between -1 and 1 in Liu Lexicon. As a result, the accumulated scores of each record from some lexicons will be hugely bigger or smaller than the others. This will cause the lexicons to have non-uniform effects on the clustering of the records. Thus, for the sake of consistency, the lexicon score matrix was normalized [50] so that the accumulated scores take decimal values between -1 and 1.

**Clustering:** The processed and consistent score matrix is utilized as the features of the data entries. Due to the large amount of textual information, it is infeasible to label the data manually. Therefore, we use clustering as an unsupervised learning method. In particular, we have employed *k*-means, which provides a high accuracy in comparison to other clustering methods such as the density based clustering or hierarchical clustering [51]. It is a partitional clustering algorithm, which in this paper functions based on the squared-error criterion. It clusters data by grouping each of the objects to the nearest cluster, based on Euclidean distances[5]. The algorithm selects the initial centroids randomly. Consequently, in a case of inappropriate selection of centroids, the partitions would be ill-suited. To avoid such a scenario, and assess the reliability of results, the code was repeated three times and we have selected the optimal number of clusters based on parameter sensitivity analysis.

To determine a suitable number of clusters for each dataset, we use the "elbow" method. The idea is to run the *k*-means clustering for a range of *k* values (i.e. *k* from 2 to 10), and for each value of *k*, the cost function is calculated for all output clusters, which is the Sum of Squared Error (SSE):

$$SSE = \sum_{i=1}^{k} \sum_{x \in C_i} (X - centroid_i)^2 \qquad (1)$$

In equation 1, *X* represents the score of each record. Next, the SSE is plotted for each number of *k*. The elbow method assumes that there will be a sudden change in the slope of the plot in a point, which means the addition of any more clusters will not result in a significant decrease of the SSE. That number will be chosen as the number of clusters for a certain dataset.

The *k*-means algorithm was applied to the four datasets using the Python "Scipy" library[6]. Having decided on the suitable number of clusters, the clustering for each set of records (Twitter, Kickstarter, Bing News, and Google News) was implemented. The centroid of each cluster was calculated and plotted for further deductions regarding public sentiments and media viewpoints towards robots and their AI applications.

---

[5] https://scikit-learn.org/stable/modules/clustering.html
[6] https://docs.scipy.org/doc/scipy/reference/cluster.vq.html

**Theme analysis:** In order to identify the robot types that correspond to the highest number of favorable or unfavorable public sentiments, a theme study on the tweets has been performed. For this study a random collection of 180 tweets, distributed equally over all clusters, has been selected. Two researchers independently annotated each tweet. Those researchers identified 10 major categories of robots and evaluate the usage of each category of robots in each of the clusters. This will provide us with information over which robot category is most discussed in which cluster.

**Kappa statistic:** In an attempt to measure the inter-rater agreement between the two raters of this study, Kappa static is employed. Here, the approach of Fleiss kappa is utilized [52]. It is an adaptation of Cohen's Kappa and is applicable for any number of raters. The kappa can be calculated using the formula below:

$$\kappa = \frac{Pr(a) - Pr(e)}{1 - Pr(e)} \quad (2)$$

Where Pr(a) denotes the agreement which is actually observed and Pr(e) stands for the chance agreement.

**Topic evolution:** Text entries are typically clustered by subject and source, but a compelling factor to consider while performing clustering on textual data can be time [53]. While it is absolutely essential to know which topics are attracting more attention and investments towards themselves at the moment, it cannot be denied that the changes in the trend of each topic over time is just as important for the involved industries and investors.

A new method of word clustering is devised in a study on topic evolution [26]. We have employed this method to assess the evolution of the social media discussion topics. The set of news and Kickstarter records of each year, from 2011 to 2018, were analyzed and targeted by this word-oriented clustering method.

## 5 RESULTS

### 5.1 Cluster Distributions

The optimal number of clusters for each dataset is decided by analyzing the scree plots illustrated in Fig. 3. The optimal numbers correspond to the longest perpendicular lines connecting the plot to the line between the first and the last point. Consequently, the optimal numbers of 4, 3, 6, and 4 are determined for datasets of Kickstarter, Bing News, Twitter, and Google News respectively. We report the cluster centroids of each dataset in Fig. 4, plotting them based on their scores from the lexicons. On the other hand, the centroids are assumed to represent the members of the clusters. Thus, the corpuses are summarized in the plots of Fig. 4. It is worth mentioning that in order to validate the clustering process Mann–Whitney–Wilcoxon (MWW) [56] test is conducted where p-value<0.5. It is observed that Kickstarter, Bing News and Google News consist of neutral and positive clusters while Twitter include one negative cluster as well. In addition, Fig. 5 illustrates the population distributions of clusters in each dataset.

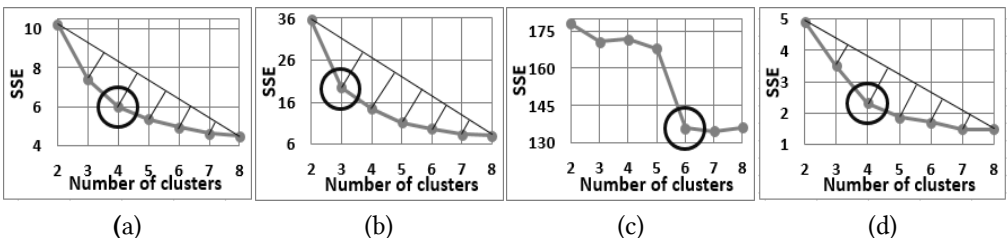

(a)  (b)  (c)  (d)

Fig. 3. Scree plot for (a) Kickstarter (b) Bing News (c) Twitter and (d) Google News

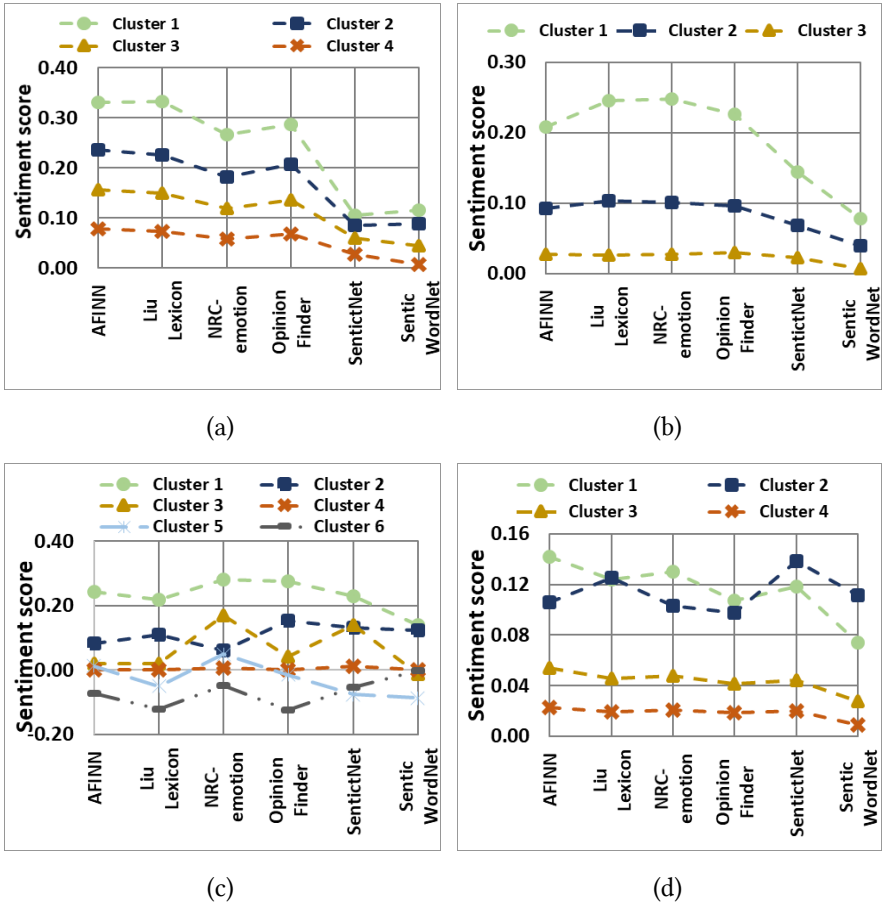

Fig. 4. Cluster distributions for (a) Kickstarter (b) Bing News (c) Twitter and (d) Google News

As for the Kickstarter dataset, four clusters have emerged. Cluster 4, which compromises the biggest part of population (38% of posts), is almost objective. All other clusters embody the positive points of view, which range from slightly positive (cluster 3 with 36% of posts) to considerably positive (cluster 1 with 5% of posts). Similarly, the most populated cluster of the Bing News dataset is the objective cluster, containing 74% of the population. The remaining two, cluster 2 with moderately positive sentiments and cluster 1 with mostly positive ones, comprise 22% and 4% of posts respectively. Tweets, as opposed to the other datasets, include one negative cluster, which includes 10% of the entire population of the dataset. Approximately half of the dataset, inclusive of cluster 4 with 43% of tweets and cluster 5 with 5% of tweets, qualify as objective. The remaining 42% percent of tweets belong to the positive opinions, of which 7% are highly polarized. Ultimately, in the Google News dataset, the objective contents (cluster 3 and 4) shape the majority. The supportive News (cluster 1 and 2) occupy close to one third of the population.

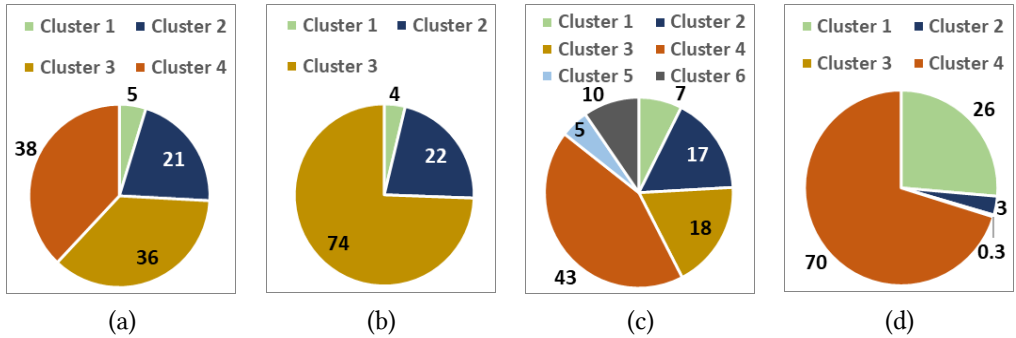

Fig. 5. Populations percentages of clusters for (a) Kickstarter (b) Bing News (c) Twitter and (d) Google News

## 3.2 Theme Analysis

Two of the researchers independently annotated 180 randomly chosen tweets and came up with 10 main categories of robots (Table 2). They are identified progressively during the annotation process of theme extraction. The annotation resulted to an almost perfect agreement with a Fleiss kappa inter-rater agreement of 83.9%.

Table 2. Robot categories identified from inductive theme analysis

| | | | |
|---|---|---|---|
| 1 | Robophobia<br>Human-robot trust<br>Robot abusing behaviors<br>Robot autonomy<br>Robot right | 4 | Educational robots |
| | | 5 | Military robots |
| | | 6 | Humanoid robots<br>Robot embodiments |
| | | 7 | Sex robots |
| 2 | Interactive robots<br>Companion robots<br>Social robots | 8 | Agriculture robots |
| | | 9 | Artificial Intelligence<br>Swarm Robotics |
| 3 | Assistive robots<br>Health care robots<br>Rescue robots<br>Collaborative robots | 10 | Advertisements |

Referring back to Fig. 4 section (c), cluster 1 to 3 represent the popular robot types as opposed to cluster 6, which embodies the negative sentiments. Table 2 reports the result of the theme extraction where it is observed that in the three positive clusters the majority belongs to robot category 3 (assistive robots), 7 (sex robots), and 2 (interactive robots), respectively. Furthermore, it is interesting that sex robots take second place. At the same time, sex robots are the dominant category in cluster 6 and has attracted negative feedbacks the most. This result reveals a major controversy in social media about sex robots. Estimating the number of their proponents and opponents shows that the former has 4,546 members, while the latter has 2,367, which is about half the size. In other words, the supporters of sex robots are approximately twice more than their opponents.

Table 3. Percentage of sentiment distribution of each cluster among the robot categories

| Robot category | 1) Robophobia | 2) Interactive robots | 3) Assistive robots | 4) Educational robots | 5) Military robots | 6) Humanoid robots | 7) Sex robots | 8) Agriculture robots | 9) AI | 10) Ads |
|---|---|---|---|---|---|---|---|---|---|---|
| Cluster 1 (very positive) | 3 | 22 | 33 | 0 | 0 | 7 | 28 | 3 | 3 | 0 |
| Cluster 2 (positive) | 5 | 3 | 27 | 0 | 0 | 30 | 30 | 0 | 5 | 0 |
| Cluster 3 (slightly positive) | 5 | 13 | 25 | 3 | 0 | 18 | 27 | 0 | 5 | 3 |
| Cluster 4 (neutral) | 7 | 7 | 30 | 0 | 2 | 8 | 3 | 0 | 3 | 40 |
| Cluster 5 (neutral) | 10 | 15 | 32 | 0 | 0 | 17 | 22 | 0 | 5 | 0 |
| Cluster 6 (negative) | 0 | 12 | 12 | 0 | 5 | 5 | 65 | 0 | 2 | 0 |

Surprisingly, assistive and interactive categories have gained almost a quarter of negative views. Non-polarized clusters (clusters 4 and 5) mostly consist of robot, workshop, and conference advertisements along with News regarding group of assistive robots.

## 5.3 Topic Evolution Analysis

The word-oriented clustering algorithm utilized in this study results in clusters consisting of the exact words used in the articles, which falls under similar categories. These categories can be "nouns", "adjectives", "industry names", "robot types", "locations", "positive or negative adverbs reflecting concerns", etc. This method emphasizes two features to differentiate between the resulting clusters: *Centrality* and *Density*. The former implies the dominant theme, around which the words of each cluster are gathered, and the later expresses the density with which these words appear throughout each set of entries (each set belongs to a certain year).

The outcome clusters can be summarized into a set of tables demonstrating main words and topics (centrality) used most (density) by users in social media, divided by each year. Comparing the tables belonging to different years can be useful in understanding the evolution of each cluster in time and understanding the rise and fall of words and topics of discussion over the news and Kickstarter articles. Taking into account the cluster densities, and the strength of words inside each cluster, all these tables can be combined to make a list of top words. These words, put into eight main categories, form Table 4 categorizing main public discussion topics.

Table 4. Public discussion main topics and sub-topics

| Main Categories | Top Words |
| --- | --- |
| Robot Efficiency | Energy, Efficient, Effective, Faster, Cheaper, Accurate, Smaller |
| Healthcare Robots | Medicine, Nursing, Surgery, Healthcare, Therapy |
| Military Robots | War, Military, Weapon, Security |
| Social Robots | Social, Privacy |
| Educational Robots | Education, School, Learning, Teaching |
| Industrial Robots | Industry, Factory, Manufacturing |
| Autonomous Robots/Drone/Car | Drone, Autonomous, Car, Service |
| Support / Service Robots | Telepresence, Exoskeleton, Humanoid, Service, Support, Wearable, Sex |

A thorough analysis of these topics and keywords over time (from 2011 to 2018), demonstrates that some topics evolve over time, while others lose importance. The results are illustrated in a heat map (Fig. 6), where strength of color shows the estimated importance of each topic throughout an entire year.

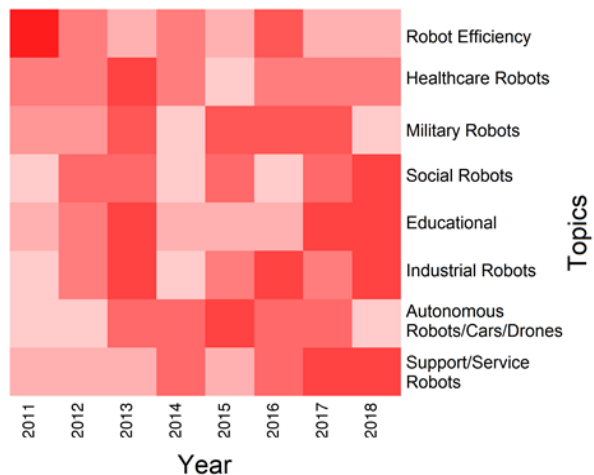

Fig. 6. Heat map of topic significance illustrated over time

As visible in the illustration, all topics both gain and lose importance over the time period during which this study was done. Due to the nature of scientific advancement, it is to be expected that articles regarding certain fields of technology will trend during time periods in which the field in question is thriving. Taking this into account, it is possible to deduce the prevailing areas of robotics or AI over each period of time throughout the 2010s. This can result in estimating the state of investments and scientific attractions, over the entire field, at the time. By the same logic, it is possible to make predictions for the future of the robotics industry, based on the trending public discussions or news articles in the recent months or years. According to the heat map, more intense colors in the recent year belong to the fields of healthcare, social, educational, service and support robots.

## 6 FINDINGS AND DISCUSSION

Here we summarize our findings from our analysis, how they address our research questions, and what are open questions worth further investigation by the community.

**Public's opinion advocates robots:** Following the application of clustering on the entirety of the four datasets (section 5.1), it is interpretable that all the datasets share one main feature: the subjective opinions take a positive position. This finding answers the first part of RQ1. The sole negative cluster is observed among tweets with 10% of the total population. Nevertheless, in the same community the positive sentiments represent nearly a quarter of the population, which is still excessively more than the negative notions.

**Media take similarly a positive position, but are more neutral than the public:** Tweets and Kickstarter posts compromise a total of 42% and 62% of positive views. In contrast, as for news (Google News and Bing News) the positivity is a mere average of 28%. News media have engaged considerably less in positive feedback. Instead, they have been more involved with non-polarized news. The non-polarized posts of news platforms are about ¾, whereas those posts are about ⅓ for Kickstarter and half for the tweets. The majority of neutral posts exhibit their conservatively. However, their non-polarized posts reveal that news agencies have not resisted against the positive trend toward robots, answering the second part of RQ1. Such a favorable climate presumably owes much to the advancements in robot performances, particularly in recent years.

**Kickstarter is the most favorable environments for robots:** In section 5.1 it is observed that the neutral posts are dominant in all sets, except for Kickstarter, where positive notions take the top place, involving 66% of the posts. This positive environment is in agreement with the nature of this platform, which is comprised of developers and small business owners.

**Sex robots are most discussed and most popular robots among the polarized tweets:** Theme analysis (section 5.2) identified that despite a fierce debate over sex robots, they have managed to be among the most popular robot types. It is observed that the proponents of sex robots are almost twice as many as the opponents. Thus, given this fierce public controversy, it would not be surprising to witness more debates in media over sex robots in the future. Furthermore, the majority of their advocates, and presumably the demands for them might trigger a significant growth in their market in the future. This immense popularity gives rise to the questions such as: 1) what is the underlying social reason(s) for embracing these robots? and 2) what are the impacts of developing these robots in society? It might turn out that these robots would lead to adverse consequences in nurturing families, and in turn result in negative social outcomes. It seems that sociologists may need to pay attention to this phenomenon, and policy makers may be required to take it into consideration proactively. This partially answers RQ2.

**There are adverse effects about social robots and assistive robots:** From the results of section 5.2, it is interpretable that after sex robots, which gain the greatest number of negative posts, the category of interactive robots and also assistive robots stand in next place, despite their popularity. The negativity concerning interactive robots can be understandable in light of their potentially adverse effect on interpersonal relationships. The unfavorable views on assistive robots might be associated with unemployment concerns. This partially answers RQ2.

**The overall conceptualization of robots has changed over recent years:** The study of the evolution of social and news media topics (section 5.3), discussed over an almost decade-long period, shows that the overall perception of the concept that surrounds robots has changed notably during this time period. Points such as the drop in the importance of particular topics, such as "Robot Efficiency" and "Autonomous Robots", and the rise in "Social Robots" and "Service Robots," showcases the shift in public and news media views on robots, from a solely industrial-purposed machine, towards a more in-home, friendly, social, and multi-purpose gadget. Today, intelligent robots have become an inseparable part of modern households, which can help one do their chores, support them in their everyday tasks, help their children study, or even monitor their health and well-being. During the mid-2010s, we witnessed the concept of electric/autonomous cars and drones were a thriving trend in the media. In reality, towards the end of the 2010s, the market share of electric cars is on a new record high [54]. The findings here are in agreement with the reality of market share rises in the present, and they answer the RQ3.

**Military robots and industrial robots are constantly key topics:** The evolution study discussed in section 5.3 reveals that, while topics such as "Military Robots" have remained to be an important concern over the past decade, the age of thriving technology has also kept "Industrial Robots" as a key topic in the field.

The years that come after, are the time for a new variety of robots. Public and media interest manifests the beginning of the "Companion Robots" era. Investors will be flooding the field with more money, and users are finding new applications for these gadgets every day. As such, it would not be surprising to witness an even bigger rise in the significance and emphasize of such discussion topics in the coming years.

There is another minor finding, which is not focused on robots. However, we identified that in comparison to other tools, SentiWordNet has exhibited a relatively poor performance. It is noticeable from section 5.1 that it has exhibited the poorest performance among all the six lexicons. In particular, it has not been capable of separating the clusters well in the Kickstarter, Bing and Twitter datasets. This lexicon works based on WORDNET synsets[7]. The synsets of WORDNET are employed in SentiWordNet, which runs a semi-supervised text classification process in order to classify the synsets into groups of positive, negative and objective [55]. This weak-supervision learning algorithm accounts for the poor capability of this lexicon observed above.

## 7 CONCLUSIONS

This paper reported the results of social research on public sentiments and news production mediums' attitudes toward robots by gathering textual records from Twitter, Kickstarter, Google News, and Bing News, and running clustering algorithms on the four datasets. The results of our evaluations revealed the public and news opinions among all platforms to be either completely, or greatly in agreement, with a minority of Twitter posts exhibiting slight concerns. In parallel, Theme Extraction and Topic Evolution methods were employed to study the textual datasets topic-wise. The two investigations expose the most discussed topics and robot types within positive and negative clusters, and the evolution of the most discussed topics, over a certain period of time. Based on these results, we discussed the present standpoint of society on the subject of robots, the evolution of that viewpoint over the past decade, and the possible outlook of a broadly robotic and AI-powered future. Ultimately, based on this evidence, we have made predictions on the subject of certain future investments in the field of robotics and AI.

---

[7] https://sentiwordnet.isti.cnr.it/